\documentclass[twocolumn,pra,superscriptaddress,showpacs,amsfonts,amssymb,amsmath,letterpaper]{revtex4}
\usepackage{amsmath}
\usepackage{graphicx}
\usepackage{amssymb}
\usepackage{pslatex}

\makeatletter
\@ifundefined{textcolor}{}
{%
 \definecolor{BLACK}{gray}{0}
 \definecolor{WHITE}{gray}{1}
 \definecolor{RED}{rgb}{1,0,0}
 \definecolor{GREEN}{rgb}{0,1,0}
 \definecolor{BLUE}{rgb}{0,0,1}
 \definecolor{CYAN}{cmyk}{1,0,0,0}
 \definecolor{MAGENTA}{cmyk}{0,1,0,0}
 \definecolor{YELLOW}{cmyk}{0,0,1,0}
 }

\usepackage{txfonts}
\usepackage{dcolumn}
\makeatother

\begin{document}

\title{
Waveguide QED: Many-Body Bound State Effects in Coherent and Fock State\\
 Scattering from a Two-Level System}

\author{Huaixiu Zheng}
\affiliation{\textit{Department of Physics, Duke University, P. O. Box 90305,
Durham, North Carolina 27708, USA}}
\affiliation{\textit{Center for Theoretical and Mathematical Sciences, Duke University,
Durham, North Carolina 27708, USA}}

\author{Daniel J. Gauthier}
\affiliation{\textit{Department of Physics, Duke University, P. O. Box 90305,
Durham, North Carolina 27708, USA}}

\author{Harold U. Baranger}

\affiliation{\textit{Department of Physics, Duke University, P. O. Box 90305,
Durham, North Carolina 27708, USA}}
\affiliation{\textit{Center for Theoretical and Mathematical Sciences, Duke University,
Durham, North Carolina 27708, USA}}

\date{September 27, 2010}

\begin{abstract}
Strong coupling between a two-level system (TLS) and bosonic modes produces
dramatic quantum optics effects. We consider a one-dimensional continuum
of bosons coupled to a single localized TLS, a system
which may be realized in a variety of plasmonic, photonic, or electronic contexts.
We present the exact many-body scattering eigenstate obtained by imposing open boundary conditions.
\emph{Multi-photon bound states} appear in the scattering of two or more
photons due to the coupling between the photons and the TLS.
Such bound states are shown to have a large effect on
scattering of both Fock and coherent state wavepackets, especially in the intermediate coupling strength regime.
We compare the statistics of the transmitted light with a coherent state
having the same mean photon number:
as the interaction strength increases, 
the one-photon probability is suppressed rapidly, and the two- and
three-photon probabilities are greatly enhanced due to the many-body bound states. This results in non-Poissonian light.
\end{abstract}


\maketitle

\section{Introduction}

Recently, there has been increasing interest in designing quantum optical
elements based on the strong coupling between light and matter
{\cite{PolitiSci08,HofheinzNat08,ChangPRL06,ChangNatPhy07,LiaoPRA09,
ZhouPRL08,ZhouPRA09,WitthautNJP10,LongoPRL10}}.
The strong coupling regime has been realized in the classic cavity quantum electrodynamics (QED) systems \cite{VogelQO03,ThompsonPRL92,ReithmaierNat04}, as well as in circuit-QED
experiments \cite{WallraffNat04,ChiorescuNat04,AstafievSci10,SchoelkopfNat08}. Several experimental systems have been proposed for realizing devices such as a single-photon transistor {\cite{ChangNatPhy07,WitthautNJP10}} or a
quantum switch {\cite{ZhouPRL08,ZhouPRA09,LongoJOA09}}, including
surface plasmons coupled to a single two-level emitter {\cite{ChangNatPhy07}}, a
superconducting transmission line resonator coupled to a local superconducting
charge qubit {\cite{ZhouPRL08,ZhouPRA09}}, and propagating photons
in a 1D waveguide coupled to a two-level system {\cite{ShenPRL07,ShenPRA07}}. Most of the theoretical work focuses on a single-photon coupled
to a local quantum system modeled as a two-level system (TLS). The key property used in the device proposals is that, if the energy
of the incident photon is tuned to be on resonance with the TLS,
the system will block the transmission of photons due to destructive
interference between the directly transmitted photon and the
photon reemitted by the impurity {\cite{ChangNatPhy07,ZhouPRL08}}.

A more challenging task is to study the two or more photon scattering
problem in such systems. The two-photon problem has been addressed
by Shen and Fan using a generalized Bethe-ansatz {\cite{ShenPRL07,ShenPRA07}}.
They showed that two-photon bound states emerge 
% as a composite particle 
as the photons interact with the two-level system. Effective
attractive and repulsive interactions can be induced depending on
the energy of the photons {\cite{ShenPRL07}}. Such effective interactions
between photons may provide new avenues for controlling photon 
entanglement \cite{MaunzNatPhys07}.
However, the scattering eigenstates were not constructed explicitly
in Ref.\,\onlinecite{ShenPRA07}: the bound states were found by first
constructing Bethe-type scattering eigenstates and then deducing
the bound states via the completeness of the basis. It is difficult
to generalize the method in Ref.\,\onlinecite{ShenPRA07} to solve
the three-photon (or more) scattering problem in which we expect
more complicated and interesting photon correlations.

\begin{figure}[t]
\centering
\includegraphics[width=3.0in,clip]{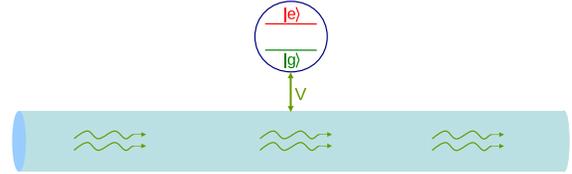}
\caption{(color online) Sketch of the structure considered: a 1D continuum of bosons coupled to a two-level-system.}
\label{fig:structure}
\end{figure}

Here, we present a method to explicitly construct exact $n$-photon
scattering eigenstates and then use the eigenstates to analyze the scattering
of Fock- and coherent-state wavepackets. The system consists
of a 1D bosonic continuum coupled to a local two-level-system as shown in Figure\,\ref{fig:structure}. First,
we explicitly construct the $n$-photon ($n=1$ to $4$) scattering eigenstates
by imposing open boundary conditions while requiring that the incoming wavefunctions consist entirely of
plane waves \cite{NishinoPRL09,ImamuraPRB09}. In addition to
two-photon bound states, three-photon bound states appear in the three-photon
scattering eigenstates, and likewise $n$-photon bound states appear in the $n$-photon scattering eigenstates. Second, to show the significance of these bound
states in the scattering of practical light sources, we study the
scattering of one-, two-, and three-photon Fock state wavepackets. It is shown
that the two- and three-photon bound states dramatically enhance the transmission of two- and three-photon
wavepackets, respectively. Third, we study the scattering of
coherent states to determine the impact of the bound states on both the photon
correlation and the statistics of the transmitted and reflected photons. Strong bunching and antibunching
effects appear, and the statistics are non-Poissonian.

The paper is organized as follows. In Sec.\,II, we introduce the model,
solve for the $n$-photon scattering eigenstates in the $n=1$ to $4$ cases,
and construct the corresponding S-matrix based on the Lippmann-Schwinger
formalism \cite{SakuraiQM94}. In Sec.\,III, the impact of bound states on the
photon transmission is studied for initial 
Fock-state wavepackets with photon number of one, two, and three. 
In Sec.\,IV, we present the analysis of photon correlation and statistics
for coherent state scattering. Finally, we conclude in Sec.\,V.

\section{Scattering Eigenstates}

The system we study consists of a two-level system coupled to photons
propagating in both directions in a one-dimensional waveguide \cite{ShenPRL07,ShenPRA07,LongoPRL10}.
The system is modeled by the Hamiltonian \cite{ShenPRL07}

\begin{eqnarray}
\lefteqn{H =\int dx\frac{1}{i}\Big[a_{R}^{\dagger}(x)\frac{d}{dx}a_{R}(x)
-a_{L}^{\dagger}(x)\frac{d}{dx}a_{L}(x)\Big]} & & \nonumber \\[6pt]
 & & +\Big(\epsilon-\frac{i\Gamma'}{2}\Big)|e\rangle \langle e|
+\int dx V \delta(x)\Big\{\big[a_{R}^{\dagger}(x)+a_{L}^{\dagger}(x)\big]S^{-}+h.c.\Big\},\;\;\quad
\label{eq:Ham}
\end{eqnarray}
where $a_{R}^{\dagger}(x)/a_{L}^{\dagger}(x)$ is the creation operator
for a right-going/left-going photon at position $x$, $\epsilon$ is the level splitting between the ground state $|g\rangle$ and the excited
state $|e\rangle$ of the two level system, $\Gamma'$
is the decay rate into channels other than the 1D continuum, 
$V$ is the frequncy-independent coupling strength, and $S^{-}=|g\left\rangle \right\langle e|$
is the atomic lowering operator. Throughout the paper, we set the group velocity
$c$ and Plank's constant $\hbar$ to $1$ for simplicity. 

It is natural to transform to modes which are either even or odd about the origin,
$a_{e/o}^{\dagger}(x)\equiv \big[a_{R}^{\dagger}(x)\pm a_{L}^{\dagger}(-x)\big]/\sqrt{2}$.
The Hamiltonian (\ref{eq:Ham}) is then decomposed into two decoupled
modes: $H=H_{e}+H_{o}$ with
\begin{subequations}
\begin{eqnarray}
H_{e} & =&\int dx\frac{1}{i}a_{e}^{\dagger}(x)\frac{d}{dx}a_{e}(x)+\left(\epsilon-i\Gamma'/2 \right)|e\rangle \langle e|\nonumber \\
 & &+\int dx\,\bar{V}\,\delta(x) \big[ a_{e}^{\dagger}(x)S^{-}+h.c.\big],\\[6pt]
H_{o}&=&\int dx\frac{1}{i}a_{o}^{\dagger}(x)\frac{d}{dx}a_{o}(x),
\end{eqnarray}
\end{subequations}
where the effective coupling strength becomes $\bar{V}=\sqrt{2}V$. Note that the odd mode is free.
The number operator for even bosons is $n_e=\int dx\,a_{e}^{\dagger}(x)a_{e}(x)$, that for the odd bosons is $n_o=\int dx\,a_{o}^{\dagger}(x)a_{o}(x)$, and the occupation number of the two-level system is $n_{tls}=|e\rangle \langle e|$.
Because $H$ commutes with certain number operators,
$[H,\,n_e+n_{tls}]=[H,\,n_o]=0$, the total number of excitations in both the even and odd spaces are separately conserved. 
We will now focus on finding the non-trivial even-mode solution 
and then transform back to the left/right representation.

A $n$-excitation state ($n=n_e+n_{tls}$) is given by 
\begin{eqnarray}
\lefteqn{|\psi_{n}\left\rangle \right. =\int dx_{1}\cdots dx_{n}\,
g_{n}(x_{1},\cdots,x_{n})
\,a_{e}^{\dagger}(x_{1})\cdots a_{e}^{\dagger}(x_{n})|0,g\rangle } & &\nonumber \\[6pt]
& &+\int dx_{1}\cdots dx_{n-1}\,e_{n}(x_{1},\cdots,x_{n-1})\,
a_{e}^{\dagger}(x_{1})\cdots a_{e}^{\dagger}(x_{n-1})|0,e\rangle,\;\;\;\;\;
\end{eqnarray}
where $|0,g\rangle$ is the zero photon state with
the atom in the ground state. From $H_{e}|\psi_{n}\left\rangle \right.=E_{n}|\psi_{n}\left\rangle \right.,$
we obtain the Schr\"{o}dinger equations
\begin{equation} \begin{split}
&\Big[\frac{1}{i}(\partial_1+\cdots+\partial_n)-E_{n} \Big]g_{n}(x_{1},\cdots,x_{n}) \\
&+\frac{\bar{V}}{n}\Big[\delta(x_{1})e_{n}(x_{2},\cdots,x_{n})+\cdots+\delta(x_{n})e_{n}(x_{1},\cdots,x_{n-1}) \Big]=0,  \\[6pt]
&\Big[\frac{1}{i}(\partial_1+\cdots+\partial_{n-1})-E_{n}+\epsilon-i\Gamma'/2  \Big]e_{n}(x_{1},\cdots,x_{n-1})\\
&\qquad\qquad\qquad\qquad\qquad\quad\;\;\;+n\bar{V}g_{n}(0,x_{1},\cdots,x_{n-1})=0, \label{eq:Scheq}
\end{split}
\end{equation}
where the eigenvalue $E_{n}=k_{1}+k_{2}+\cdots k_{n}$, and $g_{n}(x_{1},\cdots,x_{n})$ is discontinuous at $x_{i}=0,\, i=1,\cdots n$.
In all the following calculations, we set $g_{n}(0,x_{1},\cdots,x_{n-1})=[g_{n}(0^{+},x_{1},\cdots,x_{n-1})+g_{n}(0^{-},x_{1},\cdots,x_{n-1})]/2$ \cite{NishinoPRL09,ImamuraPRB09}.
The scattering eigenstates $g_{n}(x_{1},\cdots,x_{n})$ and $e_{n}(x_{1},\cdots,x_{n-1})$
are constructed by imposing the boundary condition that, in the incident
region, $g_{n}(x_{1},\cdots,x_{n})$ is free-bosonic plane wave. That
is to say, for $x_{1},\cdots,\, x_{n}<0$,
\begin{subequations}
\begin{eqnarray} \label{eq:planewave}
 g_{n}(x_{1},\cdots,x_{n})&=&\frac{1}{n!}\sum_{Q}h_{k_{1}}(x_{Q_{1}})\cdots h_{k_{n}}(x_{Q_{n}}),  \\ 
h_{k}(x)&=&\frac{1}{\sqrt{2\pi}}e^{ikx} \;.
\end{eqnarray}
\label{eq:openBC}
\end{subequations}

For $n=1$, plane-wave solutions are sufficient to satisfy Eq.\,(\ref{eq:Scheq})
with eigenenergy $E=k$: 
\begin{subequations}
\begin{eqnarray}
g_{1}(x) &=&g_{k}(x)=h_{k}(x)[\theta(-x)+\bar{t}_{k}\theta(x)],\,\,  \\[6pt]
e_{1}&=&\frac{i}{2\sqrt{\pi}V}(\bar{t}_{k}-1), \\[6pt]
\bar{t}_{k}&=&\frac{k-\epsilon+i\Gamma'/2-i\Gamma_c/2}{k-\epsilon+i\Gamma'/2+i\Gamma_c/2},\label{eq:OneSol}
\end{eqnarray}
\end{subequations}
where $\theta(x)$ is the step function and $\Gamma_c=\bar{V}^2=2V^2$ is the spontaneous emission rate from the two-level-system to the 1D continuum. 
Note that $\bar{t}_k$ is the transmission coefficient for the even problem; because the even mode is chiral, $|\bar{t}_k|=1$ when $\Gamma'=0$.

For $n=2$,
plane-wave solutions are not sufficient to satisfy Eq.\,(\ref{eq:Scheq}). As discussed by Shen and Fan \cite{ShenPRL07, ShenPRA07}, a two-photon bound state must be included to guarantee the completeness of the basis. Here, instead of extracting the bound state through a completeness check \cite{ShenPRL07, ShenPRA07}, we construct the scattering eigenstate explicitly
and find a two-photon bound state contribution to the solution, as has been done
in the open interacting resonant-level model \cite{NishinoPRL09}.
We require the two-photon solution to satisfy Eq.\,(\ref{eq:planewave}) 
in the region $x_1, x_2<0$ and solve for the solution in other regions using Eq.\,(\ref{eq:Scheq}). This method of constructing scattering eigenstates 
can be generalized to three-, four-, and even more photon cases. In the Appendix, it is shown that the two-photon eigenstate with eigenenergy $E=k_1+k_2$ is
\begin{subequations}
\begin{eqnarray}
g_{2}(x_{1},x_{2})  &=&g_{k_{1},k_{2}}(x_{1},x_{2})=\frac{1}{2!}\Big[\sum_{Q}g_{k_{1}}(x_{Q_{1}})g_{k_{2}}(x_{Q_{2}}) \nonumber \\
  &&+\sum_{PQ}B^{(2)}_{k_{P_{1}},k_{P_{2}}}(x_{Q_{1}},x_{Q_{2}})\theta(x_{Q_{1}})\Big], \\[6pt]
e_{2}(x)&=&\frac{\sqrt{2}i}{V}[g_{2}(0^{+},x)-g_{2}(0^{-},x)], \\[6pt]
B^{(2)}_{k_{P_{1}},k_{P_{2}}}(x_{Q_1},x_{Q_{2}})  &\equiv&-(\bar{t}_{k_{P_{1}}}-1)(\bar{t}_{k_{P_{2}}}-1)h_{k_{P_{1}}}(x_{Q_{2}})h_{k_{P_{2}}}(x_{Q_{2}})\nonumber\\
  &&\times e^{(-\Gamma/2-i\epsilon)|x_{Q2}-x_{Q1}|}\theta(x_{Q_{2}}-x_{Q_{1}}).
\end{eqnarray}
\label{eq:TwoSol}
\end{subequations}
Here, $P=(P_1,P_2)$ and $Q=(Q_{1},Q_{2})$ are permutations of $(1,2)$ needed to account for the bosonic symmetry of the wavefuntion, and
$\Gamma=\Gamma_c+\Gamma'$ is the total spontaneous emission rate. The two-body
bound-state term $B^{(2)}_{k_{P_{1}},k_{P_{2}}}(x_{Q1},x_{Q_{2}})\theta(x_{Q_{1}})$
is generated when there are two photons interacting with the same
two-level-system (TLS), while the TLS can only absorb one photon at one time. 
The binding strength of the two photons depends on the total spontaneous emission rate $\Gamma$. Conceptually, two photons have
two ways of going through the TLS. One way is to pass by the TLS
independently as plane waves and gain a phase factor,
which is described by the first term of $g_{2}(x_{1},x_{2})$. The
other way is to bind together and form a bound state, which is described by the second term. The formation of the bound state can be viewed as a result of stimulated emission: the first photon excites the TLS and the passing of the second photon stimulates emission of the first photon into the same right-going 
state, hence producing the bound state.

For $n=3$, a procedure similar to that used to solve the $n=2$ case yields
\begin{widetext}
\begin{subequations}
\begin{eqnarray}
\lefteqn{g_{3}(x_{1},x_{2},x_{3}) =g_{k_{1},k_{2},k_{3}}(x_{1},x_{2},x_{3})} & &\\
 & &\quad\;\;=\frac{1}{3!} \Big[\sum_{Q}g_{k_{1}}(x_{Q_{1}})g_{k_{2}}(x_{Q_{2}})g_{k_{3}}(x_{Q_{3}})
+\sum_{PQ}g_{k_{P_{1}}}(x_{Q_{1}})B^{(2)}_{k_{P_{2}},k_{P_{3}}}(x_{Q_{2}},x_{Q_{3}})\theta(x_{Q_{2}})
+\sum_{PQ}B^{(3)}_{k_{P_{1}},k_{P_{2}},k_{P_{3}}}(x_{Q_{1}},x_{Q_{2}},x_{Q_{3}})\theta(x_{Q_{1}})\Big],\nonumber\\
& &e_{3}(x_{1},x_{2})=\frac{3i}{\sqrt{2}V}[g_{3}(0^{+},x_{1},x_{2})-g_{3}(0^{-},x_{1},x_{2})],\label{eq:ThreeSol}\\
& &B^{(3)}_{k_{P_{1}},k_{P_{2}},k_{P_{3}}}(x_{Q_{1}},x_{Q_{2}},x_{Q_{3}})\equiv2(\bar{t}_{k_{P_{1}}}-1)(\bar{t}_{k_{P_{2}}}-1)(\bar{t}_{k_{P_{3}}}-1)
h_{k_{P_{1}}}(x_{Q_{2}})h_{k_{P_{2}}}(x_{Q_{3}})h_{k_{P_{3}}}(x_{Q_{3}})e^{(-\Gamma/2-i\epsilon)|x_{Q_{3}}-x_{Q_{1}}|}
\theta(x_{Q_{32}})\theta(x_{Q_{21}}),\quad
\end{eqnarray}
\end{subequations}
\end{widetext} 
where 
\begin{figure}[b]
        \center{\includegraphics[width=0.40\textwidth]
        {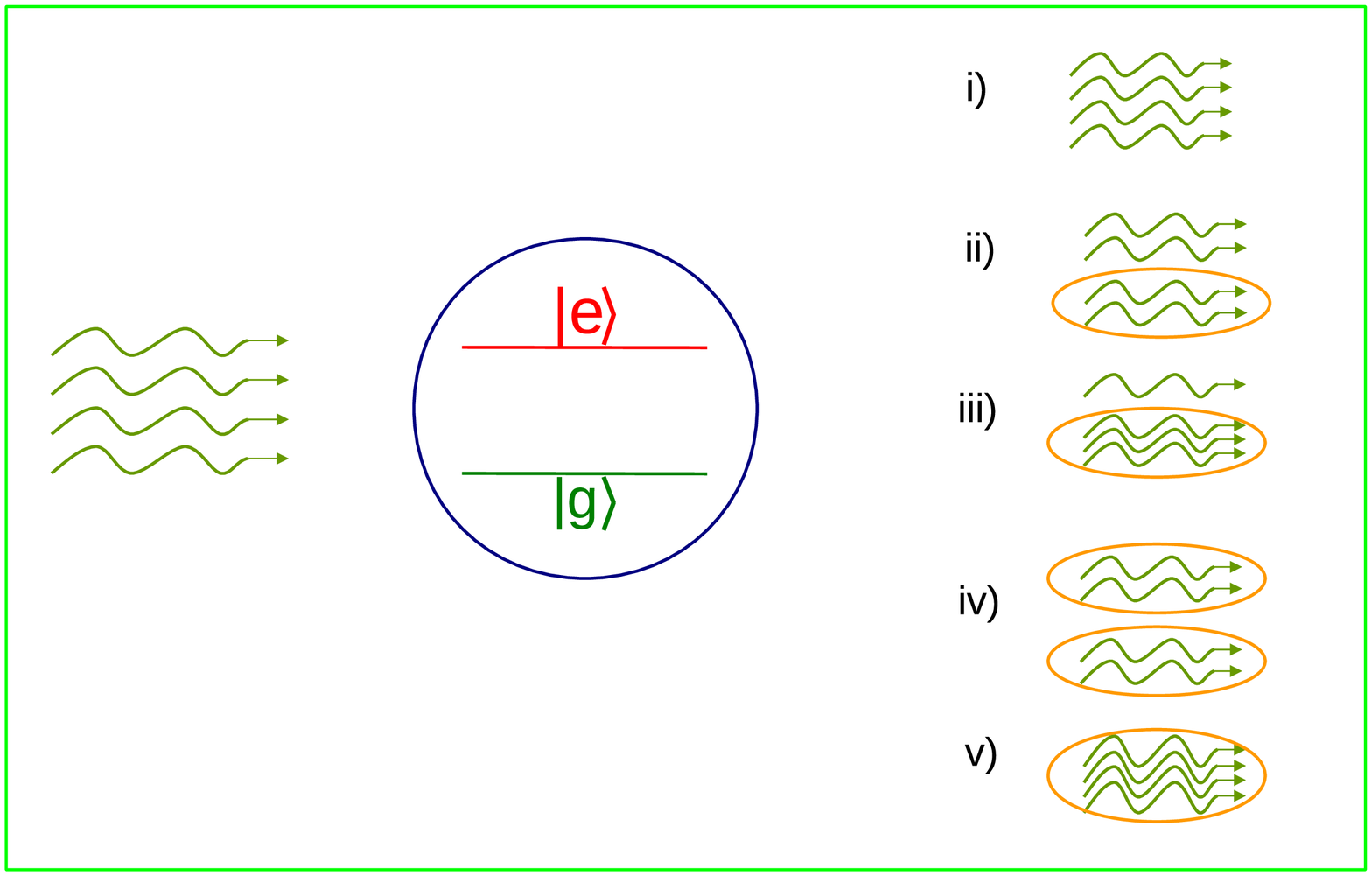}}
        \caption{\label{fig:4PSchematics} (color online) Schematic of different processes in four-photon scattering
            by a two-level system. The plane waves are represented by wiggly lines, while
            the many-body bound states are represented by the ovals.}
\end{figure}
$P=(P_{1},P_{2},P_{3})$ and $Q=(Q_{1},Q_{2},Q_{3})$ are
permutations of $(1,2,3)$ and $\theta(x_{Q_{ij}})=\theta(x_{Q_i})-\theta(x_{Q_j})$ for short. In addition to the two-photon bound state,
there emerges a three-body bound state $B^{(3)}_{k_{P_{1}},k_{P_{2}},k_{P_{3}}}(x_{Q_{1}},x_{Q_{2}},x_{Q_{3}})\theta(x_{Q_{1}})$
in the region $x_{1},x_{2},x_{3}>0$. Conceptually,
there are three ways for the three photons to pass by the atom: (i)
all three photons propagate as independent plane waves; (ii) two photons
form a two-body bound state, while the other one propagates independently
as a plane wave; and (iii) all three photons bind together and form a three-body
bound state. These three processes are described by the first, second,
and third terms of $g_{3}(x_{1},x_{2},x_{3})$, respectively.

This simple picture can be applied to a general $n$-photon scattering
process. For example, in the case of four-photon scattering, there
are five ways for the four photons to pass by the atom as illustrated in 
Figure\,\ref{fig:4PSchematics}: (i) all four
propagate as independent plane waves; (ii) two photons form a two-body
bound state, while the other two propagate independently as plane
waves; (iii) three photons form a three-body bound state, while the other
one propagate independently as a plane wave; (iv) four photons form two
independent two-body bound states; and (v) four photons form a four-body
bound state.
These five processes can be identified as the five terms
of $g_{4}(x_{1},x_{2},x_{3},x_{4})$ in the four-photon solution, which is given by
\begin{widetext}
\begin{subequations}
\begin{eqnarray}
&&g_{4}(x_{1},x_{2},x_{3},x_{4})=\frac{1}{4!}   \Big[ \sum_{Q}g_{k_{1}}(x_{Q_{1}})g_{k_{2}}(x_{Q_{2}})g_{k_{3}}(x_{Q_{3}})g_{k_{4}}(x_{Q_{4}})
+\sum_{PQ}g_{k_{P_{1}}}(x_{Q_{1}})g_{k_{P_{2}}}(x_{Q_{2}})B^{(2)}_{k_{P_{3}},k_{P_{4}}}(x_{Q_{3}},x_{Q_{4}})\theta(x_{Q_{3}}) \\
&&\qquad\qquad\qquad\qquad\; +\sum_{PQ}g_{k_{P_{1}}}(x_{Q_{1}})B^{(3)}_{k_{P_{2}},k_{P_{3}},k_{P_{4}}}(x_{Q_{2}},x_{Q_{3}},x_{Q_{4}})\theta(x_{Q_{2}})
+\sum_{PQ}B^{(2)}_{k_{P_{1}},k_{P_{2}}}(x_{Q1},x_{Q_{2}})B^{(2)}_{k_{P_{3}},k_{P_{4}}}(x_{Q_{3}},x_{Q_{4}})\theta(x_{Q_{1}})\theta(x_{Q_{3}}) \nonumber \\
&&\qquad\qquad\qquad\qquad\;+\sum_{PQ}B^{(4)}_{k_{P_{1}},k_{P_{2}},k_{P_{3}},k_{P_{4}}}(x_{Q_{1}},x_{Q_{2}},x_{Q_{3}},x_{Q_{4}})\theta(x_{Q_{1}})  \Big], \nonumber\\[6pt]
&&e_{4}(x_{1},x_{2},x_{3})=\frac{4i}{\sqrt{2}V}[g_{4}(0^{+},x_{1},x_{2},x_{3})-g_{4}(0^{-},x_{1},x_{2},x_{3})],\label{eq:FourSol}\\[6pt]
&&B^{(4)}_{k_{P_{1}},k_{P_{2}},k_{P_{3}},k_{P_{4}}}(x_{Q_{1}},x_{Q_{2}},x_{Q_{3}},x_{Q_{4}})\equiv-2^{2}(\bar{t}_{k_{P_{1}}}-1)(\bar{t}_{k_{P_{2}}}-1)
(\bar{t}_{k_{P_{3}}}-1)(\bar{t}_{k_{P_{4}}}-1)h_{k_{P_{1}}}(x_{Q_{2}})h_{k_{P_{2}}}(x_{Q_{3}})h_{k_{P_{3}}}(x_{Q_{4}})h_{k_{P_{4}}}(x_{Q_{4}}) \nonumber\\
&&\qquad\qquad\quad\qquad\qquad\quad\qquad\qquad \times e^{(-\Gamma/2-i\epsilon)|x_{Q_{4}}-x_{Q_{1}}|}\theta(x_{Q_{4}}-x_{Q_{3}})\theta(x_{Q_{3}}-x_{Q_{2}})\theta(x_{Q_{2}}-x_{Q_{1}}). 
\end{eqnarray}
\end{subequations}
\end{widetext}

The scattering eigenstates of a general
$n$-photon problem can be constructed recursively in a similar way: the only unknown term in $g_{n}(x_{1},\cdots,x_{n})$
is the $n-$photon bound state as all the other terms can be constructed
from the solutions of the $1,2,\cdots,(n-1)$-photon problems. We extrapolate from the results of $n=2$-$4$ that, for general $n$ ($\geq2$), the $n$-body bound state assumes the form
\begin{eqnarray}
\lefteqn{ B_{k_1,\dots,k_n}(x_1,\dots,x_n)=-(-2)^{n-2}\prod_{i=1}^{n}(\bar{t}_{k_i}-1)\prod_{i=1}^{n-1}\theta(x_{i+1}-x_i)} && \nonumber \\[6pt]
&&\times h_{k_1}(x_n)h_{k_2}(x_2)\cdots h_{k_{n-1}}(x_{n-1})h_{k_n}(x_n)e^{(-\Gamma/2-i\epsilon)|x_{n}-x_{1}|}.\;\;\;\;
\end{eqnarray}
We have verified this expression for $n=5$. Thus we have given explicit formulas for constructing the exact $n$-photon scattering eigenstates.

The exact scattering eigenstates can be used to construct
the scattering matrix. According to the Lippmann-Schwinger formalism
\cite{SakuraiQM94}, one can read off the ``in'' state
(before scattering) and the ``out'' state (after scattering) of
a general $n$-photon S-matrix from $g_{n}(x_{1},\cdots,x_{n})$ in
the input region ($x_{1}<0,\cdots,x_{n}<0$) and in the output region
($x_{1}>0,\cdots,x_{n}>0$), respectively. 
The ``in'' and ``out'' states of one and two photon scattering matrices are given by
\begin{subequations}
\begin{eqnarray}
&&|\phi_{in}^{(1)}\left\rangle \right._{e}=\int dx\,h_{k}(x)a_{e}^{\dagger}(x)|0\rangle  \\
&&|\phi_{out}^{(1)}\left\rangle \right._{e}=\int dx\, \bar{t}_{k}h_{k}(x)a_{e}^{\dagger}(x)|0\rangle,
\end{eqnarray}
\end{subequations}
and
\begin{eqnarray}
\lefteqn{|\phi_{in}^{(2)}\left\rangle \right._{e}=\int dx_{1}dx_{2}\frac{1}{2!} \Big[\sum_{Q}h_{k_{1}}(x_{Q_{1}})h_{k_{2}}(x_{Q_{2}}) \Big]a_{e}^{\dagger}(x_{1})a_{e}^{\dagger}(x_{2})|0\rangle } & & \nonumber \\
&&|\phi_{out}^{(2)}\left\rangle \right._{e}=\int dx_{1}dx_{2}\frac{1}{2!} \Big[\sum_{Q}\bar{t}_{k_{1}}\bar{t}_{k_{2}}h_{k_{1}}(x_{Q_{1}})h_{k_{2}}(x_{Q_{2}}) \nonumber \\
&&\qquad\qquad+\sum_{PQ}B_{k_{P_{1}},k_{P_{2}}}(x_{Q_{1}},x_{Q_{2}}) \Big]a_{e}^{\dagger}(x_{1})a_{e}^{\dagger}(x_{2})|0\rangle, \quad\quad\quad
\end{eqnarray}
and similarly for three and four photons. The corresponding S-matrices are
\begin{equation}
S_{e}^{(n)}=\int dk_{1}\cdots dk_{n}\frac{1}{n!}|\phi_{out}^{(n)}\left\rangle \right._{ee}\left.\right\langle \phi_{in}^{(n)}|.
\end{equation}
Notice that the unitarity of the S-matrix is automatically satisfied
since the incoming state $|\phi_{in}^{(n)}\left\rangle \right._{e}$ is
a complete basis set in the even space \cite{ShenPRA07,SakuraiQM94}. 

The S-matrix in the odd
space is just the identity operator because the odd mode is free and
decoupled from the impurity and the even mode,
\begin{subequations}
\begin{eqnarray}
&&S_{o}^{(n)}=\int dk_{1}\cdots dk_{n}\frac{1}{n!}|\phi_{in}^{(n)}
\rangle_{o\;o} \langle \phi_{in}^{(n)}|, \\
&&|\phi_{in}^{(n)}\left\rangle \right._{o}=\int dx_{1}\cdots dx_{n}\frac{1}{n!}\sum_{Q}\prod_{i=1}^{n}h_{k_{i}}(x_{Q_{i}})a_{e}^{\dagger}(x_{i})|0\rangle \;.
\quad\quad
\end{eqnarray}
\end{subequations}

Finally, we wish to construct the scattering matrix in the right/left
representation based on the S-matrices in the even/odd representation.
For a general $n$-photon scattering problem, the possible scattering
channels are that $i$ photons undergo scattering in the even space and
$n-i$ photons undergo scattering in the odd space, with $i$ running
from $0$ to $n$. In addition, the even and odd spaces are decoupled from each other. 
Therefore, the $n$-photon S-matrix is
\begin{equation}
S^{(n)}=\sum_{i=0}^{n}S_{e}^{(i)}\otimes S_{o}^{(n-i)}\;.\label{eq:Smatrix}
\end{equation}
We will use this S-matrix to study the scattering of Fock states and
coherent state wave packets in the right/left space in the subsequent
sections.

\section{Scattering of Fock States}

In order to show the significance of the many-body bound
states, we study the scattering of a Fock state off of a 
two-level system. We assume that the incident mode propagates
to the right and the two level system is initially in the ground state. 
We use the S-matrices defined in Eq.\,(\ref{eq:Smatrix}) to evaluate the transmission and 
reflection coefficients. In practice, any state that contains a finite number of photons must have the
form of a wave-packet.
%, since there is no physical state that spreads over the whole space.
Thus, we start with the definition of the continuous-mode photon wave-packet 
creation operator in momentum space \cite{LoudonQTL03}
\begin{equation}
 a_{\alpha}^{\dagger}=\int dk\, \alpha(k) a^{\dagger}(k)|0\rangle, 
\end{equation}
with the normalization condition $\int dk\, |\alpha(k)|^2=1$.
The corresponding continuous-mode $n$-photon Fock state is
\begin{equation}
 |n_{\alpha}\rangle=\frac{(a_{\alpha}^{\dagger})^n}{\sqrt{n!}} |0\rangle\;,
\end{equation}
and the output state after it scatters off the TLS is
\begin{equation}
 |\text{out}_{\alpha}^{(n)}\rangle=S^{(n)}|n_{\alpha}\rangle.
\end{equation}

To obtain the scattering probabilities of a Fock state from the S-matrix found in Section\,II, we follow the following general procedure. 
(i) First, we write an $n$-photon input Fock state traveling to the right in momentum space: 
$|n_{\alpha}\rangle=(1/\sqrt{n!})\int dk_1\cdots dk_n\; \alpha(k_1)\cdots \alpha(k_n)|k_1,\cdots,k_n\rangle$;
(ii) Next, we apply the S-matrix on the input state and find the output state 
$|\text{out}_{\alpha}^{(n)}\rangle=S^{(n)}|n_{\alpha}\rangle=(1/\sqrt{n!})\int dk_1\cdots dk_n\, \alpha(k_1)\cdots \alpha(k_n)S^{(n)}|k_1,\cdots,k_n\rangle$ in the even/odd basis;
(iii) We transform back to the right/left basis. Then we project the output state onto the $n$-photon (right/left-going) momentum basis 
$|k_1,\cdots,k_n\rangle_R, \cdots, |k_1,\cdots,k_i\rangle_R\otimes|k_{i+1},\cdots,k_n\rangle_L,\cdots,|k_1,\cdots,k_n\rangle_L$ and take the absolute value square
to obtain the probabilities $P(k_1,\cdots,k_n)$ of finding the output state in $|k_1,\cdots,k_n\rangle$;
(iv) Finally, we integrate $P(k_1,\cdots,k_n)$ over $k_1,\cdots,k_n$ to obtain the total transmission and reflection probabilities. Here, a right/left-going
state is defined by a positive/negative momentum, i.e., $k_1>0,\cdots,k_n>0$ for $|k_1,\cdots,k_n\rangle_R$ and $k_1<0,\cdots,k_n<0$ for $|k_1,\cdots,k_n\rangle_L$.

For convenience, we choose Gaussian type wavepackets with the 
spectral amplitude
\begin{equation}
 \alpha(k)=(2\pi \Delta^2)^{-1/4} \exp{\Big(-\frac {(k-k_0)^2} {4\Delta^2}\Big) }\;.
\end{equation}
For all of the numerical examples in this paper, we choose 
$k_0=\epsilon$: the central frequency of the wavepacket
is on resonance with the TLS, a condition which makes the interaction between the photons and the TLS strongest. 
We take the central momentum $k_0\gg \Delta$ so that 
the narrow-band condition is satisfied. In particular, we choose $\Delta=0.1$. However, we emphasize that all the conclusions we draw are independent of the choice of $\Delta$.
That is because all the transmission and reflection probabilities are functions of $\Gamma/\Delta$, where $\Gamma=2V^2$. A different choice
of $\Delta$ does not change any of the qualitative results, but merely rescales the spontaneous emission rate $\Gamma$.
 \begin{figure}[tb]
  \center{\includegraphics[width=0.48\textwidth]{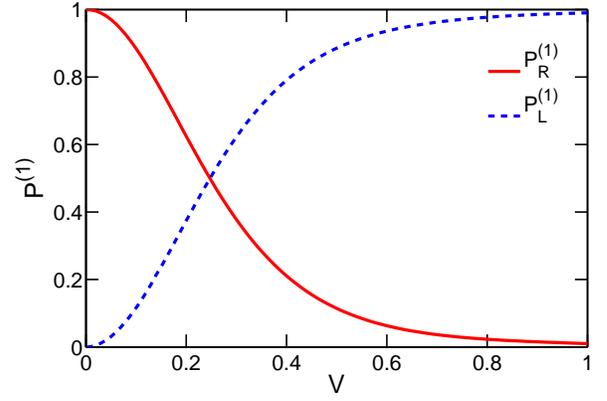}}
  \caption{\label{fig:P1} (color online) Single-photon transmission ($P^{(1)}_R$) and reflection ($P^{(1)}_L$) probabilities as a function of coupling strength $V$. The incident photon is on resonance with the two level system ($k_0=\epsilon$) and we have considered the lossless case $\Gamma'=0$.
  ($\Delta = 0.1$.) 
  }
 \end{figure}

% \begin{figure}[tbp]
% \begin{center}
% \epsfig{figure=Figs/P1.eps,width=0.45\textwidth}%\vspace{-0.4in}
% \end{center}
% %\par
% %\vspace{-0.2in}
% \caption{\label{fig:P1} Single-photon transmission ($P^{(1)}_R$) and reflection ($P^{(1)}_L$) probabilities as a function of coupling strength V.
% The incident photon is on resonance with the two level system ($k_0=\epsilon$) and we consider the lossless case $\Gamma'=0$.}
% \end{figure}
\subsection{Single-Photon Fock State Scattering}
The probabilities of transmission ($P^{(1)}_R$) and reflection ($P^{(1)}_L$) for a single-photon Fock state are found 
as
\begin{subequations}
\begin{eqnarray}
 P^{(1)}_R&=&\int_{k>0} dk \; |\langle k|\text{out}_{\alpha}^{(1)}\rangle|^2
            = \int_{k>0} dk \; \alpha(k)^2|t_k|^2,  \\
 P^{(1)}_L&=&\int_{k<0} dk \; |\langle k|\text{out}_{\alpha}^{(1)}\rangle|^2
            = \int_{k>0} dk \; \alpha(k)^2|r_k|^2,
\end{eqnarray}
\end{subequations}
where $t_k=(\bar{t}_k+1)/2$ and $r_k=(\bar{t}_k-1)/2$ and $\bar{t}_k$ is the transmission coefficient defined above for the even mode [Eq.\,(\ref{eq:OneSol})].

Note that the propagation of a single-photon is strongly modulated
by the TLS as we turn on the coupling. In the strong-coupling limit, a single-photon is perfectly reflected and the two-level atom acts as a mirror. 
This perfect reflection is due to destructive interference between the directly transmitted state and the state re-emitted from the TLS.
A single-photon transistor \cite{ChangNatPhy07} and a quantum switch \cite{ZhouPRL08} have been proposed based on this perfect reflection.

% \begin{figure}[htb]
%         \center{\includegraphics[width=0.45\textwidth]
%         {Figs/P2.eps}}
%         \caption{\label{fig:P2} Two-photon transmission and reflection probabilities as a function of coupling strength V. 
%          (a) Probability ($P^{(2)}_{RR}$) of two photon both transmitted. (b) Probability ($P^{(2)}_{RL}$) of one photon transmitted 
%              and one reflected. (c) Probability ($P^{(2)}_{LL}$) of two photon both reflected. 
%          (d) $P^2$ all together. The label PW refers to the contribution from plane wave term only, 
%             while BS refers to all the other contributions involving bound state terms.
% The incident photons are on resonance with the two level system ($k_0=\epsilon$) and we have 
% considered the lossless case $\Gamma'=0$.}
% \end{figure}

\subsection{Two-Photon Fock State Scattering}

For two incident photons, following the general procedure above, we find that
the transmission and reflection probabilities are
\begin{subequations}
\begin{eqnarray}
P^{(2)}_{RR}&=&\int_{k_1>0,k_2>0} dk_1dk_2 \;\frac{1}{2!}|\langle k_1,k_2|\text{out}_{\alpha}^{(2)}\rangle|^2,  \\
P^{(2)}_{RL}&=&\int_{k_1>0,k_2<0} dk_1dk_2 \;|\langle k_1,k_2|\text{out}_{\alpha}^{(2)}\rangle|^2,  \\
P^{(2)}_{LL}&=&\int_{k_1<0,k_2<0} dk_1dk_2 \;\frac{1}{2!}|\langle k_1,k_2|\text{out}_{\alpha}^{(2)}\rangle|^2 ,
\end{eqnarray}
\end{subequations}
where $P^{(2)}_{RR}$, $P^{(2)}_{RL}$ and $P^{(2)}_{LL}$ are, respectively, the probability for two photons to be transmitted (right-going), one 
transmitted and one reflected, and two photons reflected (left-going).

To show the significance of the bound state in the propagation of multi-photon Fock states,
we separate each of the probabilities $P^{(2)}_{RR}$, $P^{(2)}_{RL}$ and $P^{(2)}_{LL}$ into two parts. One part is
the contribution from only the plane wave term (labeled PW), which is the direct transmission or reflection. 
The other is the contribution 
from all the other terms (labeled BS), including the bound state term as well as the interference term between the plane wave and bound state.
Notice that the BS part vanishes in the absence of bound state, as in the case of single-photon scattering. Therefore, it is a manifestation of
the nonlinear effect caused by the interaction between the TLS and two or more photons. As an example, $P^{(2)}_{RR}$ split into PW and BS parts is
\begin{subequations}
\begin{eqnarray}
P^{(2)}_{RR}&=&\int_{k_1>0,k_2>0} dk_1dk_2|t(k_1,k_2)+B(k_1,k_2)|^2  \\
           &=&(P^{(2)}_{RR})_{\text{PW}}+(P^{(2)}_{RR})_{\text{BS}},  \\[6pt]
(P^{(2)}_{RR})_{\text{PW}}&=&\int_{k_1>0,k_2>0} dk_1dk_2|t(k_1,k_2)|^2,  \\[6pt]
(P^{(2)}_{RR})_{\text{BS}}&=&\int_{k_1>0,k_2>0} dk_1dk_2\big[t^*(k_1,k_2)B(k_1,k_2)  \\
&&\quad\quad +t(k_1,k_2)B^*(k_1,k_2)+|B(k_1,k_2)|^2\big] \nonumber \\[6pt]
t(k_1,k_2)&=&\alpha(k_1)\alpha(k_2)t_{k_1} t_{k_2}, \\[6pt]
B(k_1,k_2)&=&\Big[ \frac{-i/2\pi}{k_1-\epsilon+\frac{i\Gamma}{2}}+\frac{-i/2\pi}{k_2-\epsilon+\frac{i\Gamma}{2}}\Big] \\
          && \times\int_{k^{'}>0} dk^{'} \alpha(k^{'})\alpha(k_1+k_2-k^{'})r_{k^{'}}r_{k_1+k_2-k^{'}} \;. \nonumber 
\end{eqnarray}
\end{subequations}

\begin{figure}[tbp]
%\begin{center}
%\epsfig{figure=Figs/P2.eps,width=0.48\textwidth}%\vspace{-0.4in}
%\end{center}
   \center{\includegraphics[width=3.4in]{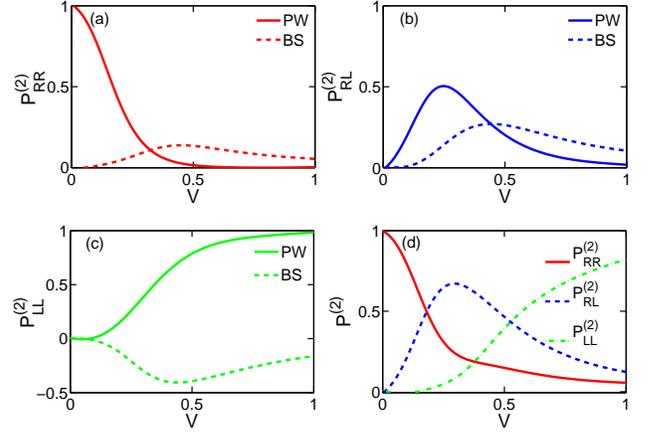}}
%\par
%\vspace{-0.2in}
\caption{\label{fig:P2} (color online) Two-photon transmission and reflection 
 probabilities as a function of coupling strength $V$. 
  (a) Probability that both photons are transmitted (and hence are right-going, $P^{(2)}_{RR}$). 
  (b) Probability that one photon is transmitted and one reflected (right-left, $P^{(2)}_{RL}$). 
  (c) Probability that both photons are reflected (both left-going, $P^{(2)}_{LL}$). 
  (d) The three processes on a single plot. 
The label PW refers to the contribution from the plane-wave term only, while BS refers to all the other contributions involving bound-state terms. 
The incident photons are on resonance with the two-level system ($k_0=\epsilon$), we consider the lossless case $\Gamma'=0$, and $\Delta=0.1$. 
Notice the large effect of the bound state on these quantities.}
\end{figure}

Figure\,\ref{fig:P2} shows the three transmission probabilities $P^{(2)}_{RR}$, $P^{(2)}_{RL}$, and $P^{(2)}_{LL}$ for our standard parameters, with the contributions from the plane wave and bound state plotted separately in panels (a)-(c). \textit{Note that the presence of the bound state has a very substantial effect on these transmission probabilities}.
As shown in panels (a) and (b), $P^{(2)}_{RR}$ and $P^{(2)}_{RL}$ are enhanced by the formation of the bound state. This is mainly due to constructive interference between the plane wave and bound state. In contrast, panel (c) shows that $P^{(2)}_{LL}$ is strongly reduced in the presence of the bound state because of destructive interference between the plane wave and bound state (change from $\sim0.8$ to $\sim0.4$ at $V=0.5$). Therefore, the presence of the bound state tends to \textit{increase} the one-photon and two-photon transmission, while \textit{suppressing} the two-photon reflection. 

A particularly interesting aspect of the results in Fig\,\ref{fig:P2} is that the effect of the bound state is most prominent in the \textit{intermediate} coupling regime, not at the strongest coupling. This is because, first, in the weak coupling limit, the interaction is too weak to produce a pronounced bound state for two-photon scattering, while, second, in the strong coupling limit, the TLS responds to the first photon too quickly (in a duration of order $1/\Gamma$ with $\Gamma=2V^2$) for the second photon to produce a significant nonlinear effect. (The formation of the bound state requires the presence of both photons at the two-level system.) The optimal coupling strength $V_m$ for producing nonlinear (bound state) effects lies at intermediate coupling, when the spontaneous emission rate $\Gamma$ is on the order of the wavepacket width $\Delta$ ($V_m\sim0.4$ when $\Delta=0.1$).

\begin{figure*}[tb]
\center{\includegraphics[width=0.98\textwidth]{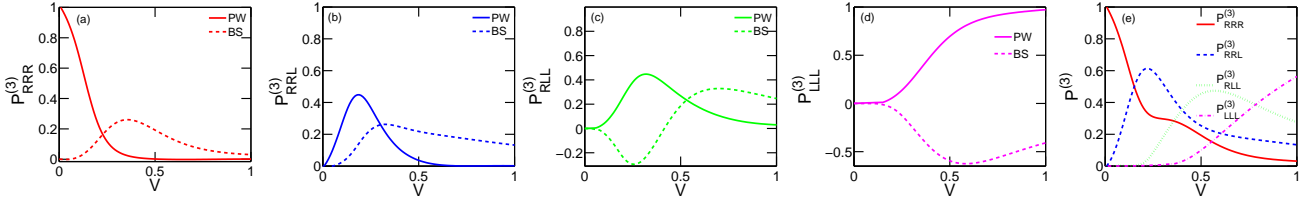}}
  \caption{\label{fig:P3} (color online) Three-photon transmission and reflection probabilities as a function of coupling strength V. 
  (a) Probability of all three photons transmitted ($P^{(3)}_{RRR}$). 
  (b) Probability of two photons transmitted and one reflected ($P^{(3)}_{RRL}$). 
  (c) Probability of one photon transmitted and two photons reflected ($P^{(3)}_{RLL}$). 
  (d) Probability of all three photons reflected ($P^{(3)}_{LLL}$). 
  (e) $P^3$ all together. The label PW refers to the contribution from only the plane wave term, while BS refers to all the other contributions, involving bound state terms. The incident photons are on resonance with the two level system ($k_0=\epsilon$), we consider the lossless case $\Gamma'=0$, and $\Delta=0.1$. Note the large bound state effects.
  }
\end{figure*}

\subsection{Three-Photon Fock State Scattering}
Following the general procedure for obtaining scattering probabilities, the transmission and reflection probabilities for three-photon Fock state scattering are defined as

\begin{eqnarray}
P^{(3)}_{RRR}&=&\int_{k_1>0,k_2>0,k_3>0} dk_1dk_2dk_3 \;\frac{1}{3!}|\langle k_1,k_2,k_3|\text{out}_{\alpha}^{(3)}\rangle|^2, \nonumber \\
P^{(3)}_{RRL}&=&\int_{k_1>0,k_2>0,k_3<0} dk_1dk_2dk_3 \;\frac{1}{2!}|\langle k_1,k_2,k_3|\text{out}_{\alpha}^{(3)}\rangle|^2, \nonumber \\
P^{(3)}_{RLL}&=&\int_{k_1>0,k_2<0,k_3<0} dk_1dk_2dk_3 \;\frac{1}{2!}|\langle k_1,k_2,k_3|\text{out}_{\alpha}^{(3)}\rangle|^2, \nonumber \\
P^{(3)}_{LLL}&=&\int_{k_1<0,k_2<0,k_3<0} dk_1dk_2dk_3 \;\frac{1}{3!}|\langle k_1,k_2,k_3|\text{out}_{\alpha}^{(3)}\rangle|^2, \nonumber \\
\end{eqnarray}
where $P^{(3)}_{RRR}$, $P^{(3)}_{RRL}$, $P^{(3)}_{RLL}$, and $P^{(3)}_{LLL}$ are the probabilities for three photons being transmitted (all right-going), two transmitted and one reflected, one transmitted and two reflected, and all three reflected (left-going), respectively. As in the two-photon scattering case, we separate each probability into two parts: the contribution of only the plane wave term (labeled PW) and the contribution from all the other terms (labeled BS), including the bound states as well as the interference between the plane wave and bound states. The probabilities and the decomposition into PW and BS parts are plotted in Figure\,\ref{fig:P3} for our usual parameters.  

Figure\,\ref{fig:P3} shows that the bound state contribution to the transmission probabilities is, as for two photons, very substantial. In panels (a) and (b), the BS parts of $P^{(3)}_{RRR}$ and $P^{(3)}_{RRL}$ are positive; thus, these probabilities are enhanced by the bound states. Panel (d) shows that $P^{(3)}_{LLL}$ is suppressed by the bound state contribution for arbitrary coupling strength. In contrast, as we increase the coupling strength, $P^{(3)}_{RLL}$ is first suppressed and then enhanced by the BS part as shown in Figure\,\ref{fig:P3}(c). Tuning the coupling strength changes the relative phase between the plane wave and bound state parts; for $P^{(3)}_{RLL}$, the interference between them happens to change from destructive to constructive as the coupling strength increases. Finally, as in the two-photon case, the most pronounced bound state effects occur in the intermediate coupling regime instead of the strong coupling limit.

To sum up this section, we point out that all the curves plotted in Figs.\,\ref{fig:P1}-\ref{fig:P3} are universal in terms of the choice of $\Delta$. Because $\Delta$ appears in the scattering probabilities [$P^{(1)}_R$, etc.] only in the ratio $\Gamma/\Delta$, a different choice of $\Delta$ (i.e., other than $0.1$ used in the figures) is equivalent to 
rescaling $V$ and does not change the shape of the curves. Therefore, the substantial bound state effects observed here are intrinsic for multi-photon scattering
processes in this system, independent of the details of the wavepackets.

\section{Scattering of coherent-states}
We now turn to studying the scattering of coherent states in order to show, first, the strong photon-photon correlation induced by the the two-level system and, second, the change in photon number statistics. The incident coherent state wavepacket is defined by \cite{LoudonQTL03}
\begin{eqnarray}
&&|\alpha\rangle=e^{a_{\alpha}^{\dagger}-\bar{n}/2}|0\rangle ,
\end{eqnarray}
with $a_{\alpha}^{\dagger}=\int dk\, \alpha(k) a^{\dagger}(k)|0\rangle$, and mean photon number $\bar{n}=\int dk|\alpha(k)|^2$.  
A Gaussian type wavepacket is chosen
\begin{equation}
 \alpha(k)=\frac{\sqrt{\bar{n}}}{(2\pi\Delta^2)^{1/4}} \exp{\Big(-\frac {(k-k_0)^2} {4\Delta^2}\Big) };
\end{equation}
for numerical evaluations, we use, as before, $\Delta=0.1$ and $k_0=\epsilon\gg\Delta$.
The output state $|\text{out}_{\alpha}\rangle$ is then
\begin{equation}
 |\text{out}_{\alpha}\rangle=\sum_n S^{(n)}|\alpha\rangle
\end{equation}
We assume the incident coherent state is right-going and the two-level system is in the ground state initially. 
We present the analysis of second-order correlation and photon number statistics in the transmitted field.

\subsection{Correlation}
The second-order correlation function of the transmitted field is defined as \cite{LoudonQTL03}
\begin{equation}
 g_{R}^{(2)}(x_2-x_1)=\frac{\langle \text{out}_{\alpha}|a_{R}^{\dagger}(x_1)a_{R}^{\dagger}(x_2) a_{R}(x_2)a_{R}(x_1)   |\text{out}_{\alpha}  \rangle}
                     {\langle \text{out}_{\alpha}|a_{R}^{\dagger}(x_1)a_{R}(x_1)   |\text{out}_{\alpha}  \rangle^2}.
\end{equation}
We consider the mean photon number $\bar{n}\leq1.0$. In this case, the probability to find $n\geq3$ number states is much smaller than that of $n=2$ number states. Moreover, the contributions from $n\geq3$ states to $g^{(2)}$ are at least one order of $\Delta$ ($=0.1$) smaller than that from the $n=2$ state.
Therefore, we neglect the contributions from $n\geq3$ number states. The second-order correlation
function simplifies to
\begin{equation}
  g_{R}^{(2)}(x_2-x_1)=\frac{\big|\int dk_1dk_2\, \alpha(k_1)\alpha(k_2)(t_{k_1}t_{k_2}-r_{k_1}r_{k_2}e^{-\frac{\Gamma(x_2-x_1)}{2}})\big|^2}
                     {\big|\int dk_1dk_2\, \alpha(k_1)\alpha(k_2)t_{k_1}t_{k_2}\big|^2}.\label{eq:g2}
\end{equation}

\begin{figure}[tb]
        \center{\includegraphics[width=0.48\textwidth]
        {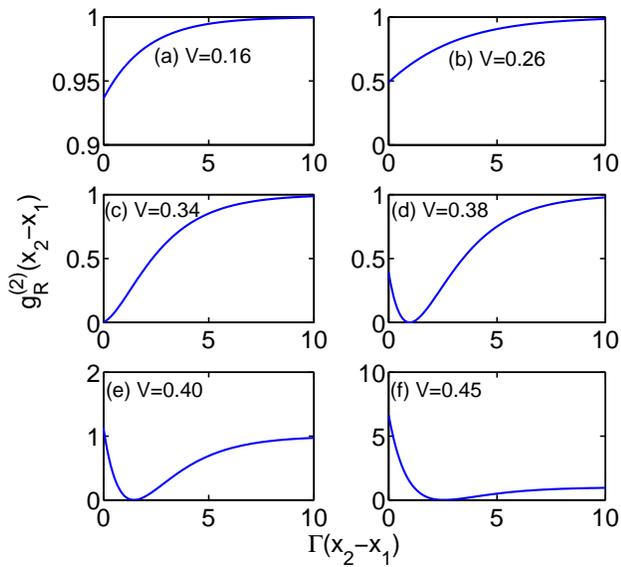}}
        \caption{\label{fig:g2} (color online) Second-order correlation of the transmitted
           field given an incident coherent state with $\bar{n}\leq1$ at various coupling strengths $V$ to the 1D continuum. (a) $V=0.16$, (b) $V=0.26$,
(c)~$V=0.34$, (d) $V=0.38$, (e) $V=0.40$, (f) $V=0.45$. The 
spontaneous emission rate  to channels other than the 1D continuum is set to $\Gamma'=0.10$. Notice that the correlation behavior is very sensitive to the coupling strength to 1D continuum, showing both bunching and antibunching.}
\end{figure}

The contributions from the directly transmitted state and the bound state can be identified as the first term and second term in the numerator of $g_{R}^{(2)}(x_2-x_1)$ in Eq.\,(\ref{eq:g2}). 
In the absence of the bound state, $g_{R}^{(2)}(x_2-x_1)$ is always equal to unity.  As we turn on the interaction, the interference 
between the directly transmitted state and the bound state will give rise to interesting correlation behavior. 
Figure\,\ref{fig:g2} shows the second-order correlation as a function of $\Gamma(x_2-x_1)$ at various coupling strengths, $V$, to the 1D mode with $\Gamma'=0.1$. 
In the weak coupling limit ($V=0.16$) as shown in Figure\,\ref{fig:g2}(a),
the directly transmitted state dominates and $g_{R}^{(2)}(0)$ is slightly smaller than 1. We observe a slight initial
antibunching. As $V$ increases [Figure\,\ref{fig:g2}(b)-(c)], $g_{R}^{(2)}(0)$ further decreases and
the initial antibunching gets stronger and becomes strongest at $V=0.34$ when $g_{R}^{(2)}(0)=0$. 
Notice that the antibunching is getting weaker as one moves away from 
the origin for $V\leq0.34$. Further increase of $V$
starts to change the initial antibunching [$V=0.38$, $g_{R}^{(2)}(0)<1$] to bunching [$V=0.45$, $g_{R}^{(2)}(0)>1$] as shown in Figure\,\ref{fig:g2}(d)-(f). 
In this case, the bound state starts to dominate the correlation behavior. It is
remarkable that, for $V>0.34$, the initial antibunching ($V<0.40$) or bunching ($V>0.40$) is followed by a later 
antibunching $g_{R}^{(2)}(0)=0$, which is caused by the cancellation of the directly transmitted state and the bound state. 
The formation of the bound state gives rise to a rich phenomenon of photon-photon correlation, which is very sensitive to the coupling strength $V$ to the 1D mode. 
\textit{Effective attractive or repulsive interaction between photons is induced by the presence of a single two-level system} \cite{ShenPRL07}. 

Our findings agree with the results obtained by Chang \textit{et al.} \cite{ChangNatPhy07} using a very different approach. 
In the lossless $\Gamma'=0$ case, as we increase the coupling strength, the transmission for individual photons is reduced
rapidly [see, for example, Figure\,\ref{fig:P1} and Figure\,\ref{fig:P2}(a)]. But the two-photon bound state can strongly enhance the transmission.
Therefore, we will observe a strong initial bunching followed by a later antibunching, similar to Figure\,\ref{fig:g2}(f).

\subsection{Photon Number Distribution}
Given the output state $|\text{out}_{\alpha}\rangle$, we measure the photon
number distribution in the transmitted field following the general procedure described in Sec.\,III.
\begin{eqnarray}
 P_0&=&|\langle\text{out}_{\alpha} |(|0\rangle_{R}\otimes |I\rangle_L) |^2, \nonumber \\
 P_1&=&\int_{k>0} dk|\langle\text{out}_{\alpha}| (|k\rangle_{R}\otimes |I\rangle_L) |^2, \nonumber \\
 P_2&=&\int_{k_1,k_2>0} dk_1dk_2\frac{1}{2!}|\langle\text{out}_{\alpha} |(|k_1,k_2\rangle_{R}\otimes |I\rangle_L) |^2, \nonumber \\
 P_3&=&\int_{k_1,k_2,k_3>0} dk_1dk_2dk_3\frac{1}{3!}|\langle\text{out}_{\alpha}| (|k_1,k_2,k_3\rangle_{R}\otimes |I\rangle_L) |^2, \nonumber\\
\end{eqnarray}
where $|I\rangle_L$ is the complete basis set in the left-going photon space. We consider a mean photon
number $\bar{n}\leq1.0$ in the incident coherent state. In this case, 
the probability to find the four photon state is negligible ($\leq1.6\%$). We compare the photon
number distribution $P_n$ of the output state with $(P_n)_{\text{Poisson}}$ of a coherent state having the
same mean photon number.

\begin{figure}[tb]
        \center{\includegraphics[width=0.48\textwidth]
        {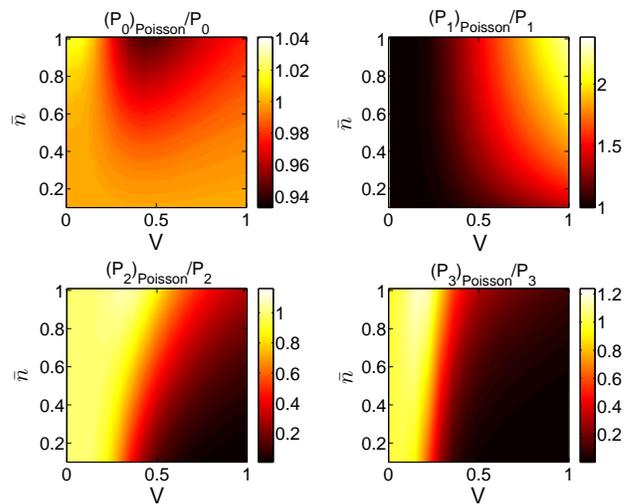}}
        \caption{\label{fig:cohstate} (color online) Photon number distribution of the transmitted
                   field compared with a coherent state. We considered the lossless case $\Gamma'=0$. The statistics is non-Possonian with the $2$ and $3$ photon content enhanced.}
\end{figure}

Figure\,\ref{fig:cohstate} shows the ratio between $(P_n)_\text{Poisson}$ and $P_n$ as a function of the coupling strength $V$ and the mean photon number $\bar{n}$ of the incident coherent state. The zero-photon probability does not deviate from that of a coherent state much in the whole parameter region we considered. The one-photon probability is smaller than the corresponding probability in a coherent state. In contrast, the two- and three-photon probabilities are much larger than the ones in a coherent state, especially in the strong coupling regime. This is to say, the interaction between photons and the two-level system redistributes the probabilities among different photon numbers. \textit{The one-photon probability is reduced and is redistributed to the two- and three-photon probabilities.} This is mainly because the bound states enhance the transmission of multi-photon states as we have shown in Sec.\,III B and C. In conclusion, we obtain a non-Poissonian light source after the scattering. It is perhaps possible to use this strongly-correlated light source to perform passive decoy-state quantum key distribution in order to raise the key generation rate \cite{HwangPRL03,LoPRL05,WangPRL05,MCOPTLETT09}.
%Coherent state definition in the form of wave-packets

%second-order correlation of the output state

%photon statistics of the output state: non-Poissonian, act as a
%filter for one-photon

%\begin{figure}[htb]
%        \center{\includegraphics[width=0.5\textwidth]
%        {Figs/P0123n1v08.eps}}
%        \caption{\label{fig:statistics} Non-Poissonian Statistics}
%\end{figure}
\section{Conclusion}
In this paper, we present a general method to construct the exact scattering
eigenstates for the problem of $n$-photons interacting with a two-level system.
Many-body bound states appear in the presence of the coupling between photons and
the two-level system. Furthermore, the scattering matrices are extracted using the Lippmann-Schwinger formalism. We emphasize 
that the completeness of the S-matrices is guaranteed by imposing open boundary
conditions and requiring the incident field to be free plane waves. Based on the S-matrices,
we study the scattering of the Fock states and coherent states. The bound states
are shown to enhance the transmission of multi-photon states and suppress the
transmission of single-photon states. In the transmitted field of coherent state scattering,
the photons exhibit strong bunching or antibunching effects depending on the
coupling strength. This is a manifestation of the many-body bound states. Finally,
we determine the photon number distribution and find that the one-photon state is
transferred to two- and three-photon states. This results in a non-Poissonian light source which might have applications in quantum information.
  
\section*{Acknowledgments}

We thank H.\,Carmichael for valuable discussions.
The work of HUB was supported in part by the U.S.\,Office of Naval Research.
HZ acknowledges support from the Graduate Program in Nanoscience and the Center for Theoretical and Mathematical Sciences, both of Duke University.

\renewcommand{\theequation}{A\arabic{equation}}
  % redefine the command that creates the equation no.
\setcounter{equation}{0}  % reset counter 
\section*{Appendix: Two-photon Scattering Eigenstate}
In this appendix, we show in detail how we obtain the two-photon scattering eigenstate [Eq.\,(\ref{eq:TwoSol})] 
by imposing the open boundary condition Eq.\,(\ref{eq:openBC}). The equations of motion for the two-photon case read
\begin{subequations}
\begin{eqnarray}
&&\Big[\frac{1}{i}(\partial_1+\partial_2)-E_{2} \Big]g_{2}(x_{1},x_{2})+ \nonumber \\
&&\qquad\qquad \frac{\bar{V}}{2}\Big[\delta(x_{1})e_{2}(x_{2})+\delta(x_{2})e_{2}(x_{1}) \Big]=0,  \\[6pt]
&&\Big[\frac{1}{i}\frac{d}{dx}-E_{2}+\epsilon-i\Gamma'/2  \Big]e_{2}(x)
+2\bar{V}g_{2}(0,x)=0, 
\end{eqnarray}
\end{subequations}
which can be cast into the following set of equations
\begin{subequations}
 \begin{eqnarray}
 &&\Big[\frac{1}{i}(\partial_1+\partial_2)-E_{2} \Big]g_{2}(x_{1},x_{2})=0,\\[6pt]
 && e_2(x)=\frac{2i}{\bar{V}}[g_2(0^+,x)-g_2(0^-,x)], \\[6pt]
  &&\Big[\frac{1}{i}\frac{d}{dx}-E_{2}+\epsilon-i\Gamma'/2  \Big]e_{2}(x) \nonumber \\
  &&\qquad \qquad +\bar{V}[g_{2}(0^+,x)+g_2(0^-,x)]=0,\\[6pt]
  &&e_2(0^+)=e_2(0^-).
 \end{eqnarray}
\end{subequations}
Here, $g_2(x_1,x_2)$ is discontinuous at $x_1=0, x_2=0$ and we set $g_2(x,0)=[g_2(x,0^+)+g_2(x,0^-)]/2$. 
We eliminate $e_2(x)$ from the above equations and obtain
\begin{subequations}
 \begin{eqnarray}
 \lefteqn{  \Big[\frac{1}{i}(\partial_1+\partial_2)-E_{2} \Big]g_{2}(x_{1},x_{2})=0,}\label{eq:P1}\\[6pt]
 \lefteqn{\Big[\frac{1}{i}\frac{d}{dx}-E_{2}+\epsilon-i\Gamma'/2-i\Gamma_c/2  \Big]g_2(0^+,x) }&&\nonumber \\
   &&\quad =\Big[\frac{1}{i}\frac{d}{dx}-E_{2}+\epsilon-i\Gamma'/2+i\Gamma_c/2  \Big]g_2(0^-,x),\label{eq:P2} \\[6pt]
   \lefteqn{g_2(0^+,0^+)-g_2(0^-,0^+)=g_2(0^-,0^+)-g_2(0^-,0^-).} \label{eq:Continuity}
\end{eqnarray}
\end{subequations}
Because of the bosonic symmetry, we can solve for $g_2(x_1,x_2)$ by first considering the half space $x_1\leq x_2$ and then extending the result
to the full sapce. In this case, there are three quadrants in real space: \textcircled{\footnotesize1} $x_1\leq x_2 <0$; \textcircled{\footnotesize2} $x_1<0<x_2$;
\textcircled{\footnotesize3} $0<x_1\leq x_2$. Eq.\,(\ref{eq:P2}) can be rewritten as two separate equations
\begin{subequations}
 \begin{eqnarray}
\lefteqn{\Big[\frac{1}{i}\frac{d}{dx}-E_{2}+\epsilon-i\Gamma'/2-i\Gamma_c/2  \Big]g_{2}^{\text{\textcircled{\tiny 2}}}(x,0^+)=} && \nonumber \\
&& \Big[\frac{1}{i}\frac{d}{dx}-E_{2}+\epsilon-i\Gamma'/2+i\Gamma_c/2  \Big]g_{2}^{\text{\textcircled{\tiny 1}}}(x,0^-),\,\, \text{for}\,\, x<0,\;\;\;\;\;\;\;\;\;\label{eq:Q1} \\[6pt]
\lefteqn{\Big[\frac{1}{i}\frac{d}{dx}-E_{2}+\epsilon-i\Gamma'/2-i\Gamma_c/2  \Big]g_{2}^{\text{\textcircled{\tiny 3}}}(0^+,x)= }&&\nonumber \\
&&\Big[\frac{1}{i}\frac{d}{dx}-E_{2}+\epsilon-i\Gamma'/2+i\Gamma_c/2  \Big]g_{2}^{\text{\textcircled{\tiny 2}}}(0^-,x),\,\,\text{for}\,\,x>0.\;\;\;\;\;\;\;\;\;\label{eq:Q2}
 \end{eqnarray}
\end{subequations}
Substituting $g_2^{\text{\textcircled{\tiny 1}}}(x_1,0^-)$ [Eq.\,(\ref{eq:openBC})] into Eq.\,(\ref{eq:Q1}), we solve to find
\begin{equation}
 g_2^{\text{\textcircled{\tiny 2}}}(x,0^+)=\frac{1}{2!}\Big[\bar{t}_{k_2}\frac{e^{ik_1x}}{2\pi}+\bar{t}_{k_1}\frac{e^{ik_2x}}{2\pi}\Big]+Ae^{[-\Gamma/2+i(k_1+k_2-\epsilon)]x},
\end{equation}
where $A$ is a constant to be determined. Applying the constraint Eq.\,(\ref{eq:P1}) to $g_2^{\text{\textcircled{\tiny 2}}}(x,0^+)$, we obtain
\begin{eqnarray}
 g_2^{\text{\textcircled{\tiny 2}}}(x_1,x_2)=\frac{1}{2!}\Big[\bar{t}_{k_2}\frac{e^{i(k_1x_1+k_2x_2)}}{2\pi}+\bar{t}_{k_1}\frac{e^{i(k_2x_1+k_1x_2)}}{2\pi}\Big] \nonumber \\[6pt]
+Ae^{(\Gamma/2+i\epsilon)(x_2-x_1)}e^{i(k_1+k_2)x_1}.\label{eq:A6}
\end{eqnarray}
From Eq.\,(\ref{eq:A6}), we can identify $A$ to be zero: otherwise, the solution is not normalizable [$e^{\Gamma(x_2-x_1)/2}$ is divergent when $x_2-x_1\rightarrow\infty$].
Hence, $g_2(x_1,x_2)$ in region \textcircled{\footnotesize2} is given by
\begin{equation}
 g_2^{\text{\textcircled{\tiny2}}}(x_1,x_2)=\frac{1}{2!}\Big[\bar{t}_{k_2}\frac{e^{i(k_1x_1+k_2x_2)}}{2\pi}+\bar{t}_{k_1}\frac{e^{i(k_2x_1+k_1x_2)}}{2\pi}\Big].\label{eq:g22}
\end{equation}
Substituting Eq.\,(\ref{eq:g22}) into Eq.\,(\ref{eq:Q2}) yields
\begin{equation}
 g_2^{\text{\textcircled{\tiny3}}}(0^+,x)=\frac{1}{2!}\bar{t}_{k_1}\bar{t}_{k_2}\Big[\frac{e^{ik_2x}}{2\pi}+\frac{e^{ik_1x}}{2\pi}\Big]+Be^{[-\Gamma/2+i(k_1+k_2-\epsilon)]x},
\end{equation}
where $B$ is a constant to be determined. Again, applying the constraint Eq.\,(\ref{eq:P1}) to $g_2^{\text{\textcircled{\tiny3}}}(0^+,x)$, we obtain
\begin{eqnarray}
 g_2^{\text{\textcircled{\tiny3}}}(x_1,x_2)=\frac{1}{2!}\bar{t}_{k_1}\bar{t}_{k_2}\Big[\frac{e^{i(k_1x_1+k_2x_2)}}{2\pi}+\frac{e^{i(k_1x_2+k_2x_1)}}{2\pi}\Big] \nonumber \\[6pt]
+Be^{(-\Gamma/2-i\epsilon)(x_2-x_1)}e^{i(k_1+k_2)x_2}.\label{eq:g23}
\end{eqnarray}
Finally, $B$ is found by substituting Eq.\,(\ref{eq:openBC}), Eq.\,(\ref{eq:g22}), and Eq.\,(\ref{eq:g23}) 
into the continuity condition Eq.\,(\ref{eq:Continuity}), yielding
\begin{equation}
 B=-\frac{(\bar{t}_{k_1}-1)(\bar{t}_{k_2}-1)}{2\pi}.
\end{equation}
Extending these solutions from the half space to the full space using the bosonic symmetry gives rise to the two-photon scattering eigenstate given in Eq.\,(\ref{eq:TwoSol}) of the main text.

%\bibliography{bspaper}
\bibliography{WQED}

\begin{thebibliography}{28}
\expandafter\ifx\csname natexlab\endcsname\relax\def\natexlab#1{#1}\fi
\expandafter\ifx\csname bibnamefont\endcsname\relax
  \def\bibnamefont#1{#1}\fi
\expandafter\ifx\csname bibfnamefont\endcsname\relax
  \def\bibfnamefont#1{#1}\fi
\expandafter\ifx\csname citenamefont\endcsname\relax
  \def\citenamefont#1{#1}\fi
\expandafter\ifx\csname url\endcsname\relax
  \def\url#1{\texttt{#1}}\fi
\expandafter\ifx\csname urlprefix\endcsname\relax\def\urlprefix{URL }\fi
\providecommand{\bibinfo}[2]{#2}
\providecommand{\eprint}[2][]{\url{#2}}

\bibitem[{\citenamefont{Politi et~al.}(2008)\citenamefont{Politi, Cryan,
  Rarity, Yu, and O'Brien}}]{PolitiSci08}
\bibinfo{author}{\bibfnamefont{A.}~\bibnamefont{Politi}},
  \bibinfo{author}{\bibfnamefont{M.~J.} \bibnamefont{Cryan}},
  \bibinfo{author}{\bibfnamefont{J.~G.} \bibnamefont{Rarity}},
  \bibinfo{author}{\bibfnamefont{S.}~\bibnamefont{Yu}}, \bibnamefont{and}
  \bibinfo{author}{\bibfnamefont{J.~L.} \bibnamefont{O'Brien}},
  \bibinfo{journal}{Science} \textbf{\bibinfo{volume}{320}},
  \bibinfo{pages}{646} (\bibinfo{year}{2008}).

\bibitem[{\citenamefont{Hofheinz et~al.}(2008)\citenamefont{Hofheinz, Weig,
  Ansmann, Bialczak, Lucero, Neeley, O'Connell, Wang, Martinis, and
  Cleland}}]{HofheinzNat08}
\bibinfo{author}{\bibfnamefont{M.}~\bibnamefont{Hofheinz}},
  \bibinfo{author}{\bibfnamefont{E.~M.} \bibnamefont{Weig}},
  \bibinfo{author}{\bibfnamefont{M.}~\bibnamefont{Ansmann}},
  \bibinfo{author}{\bibfnamefont{R.~C.} \bibnamefont{Bialczak}},
  \bibinfo{author}{\bibfnamefont{E.}~\bibnamefont{Lucero}},
  \bibinfo{author}{\bibfnamefont{M.}~\bibnamefont{Neeley}},
  \bibinfo{author}{\bibfnamefont{A.~D.} \bibnamefont{O'Connell}},
  \bibinfo{author}{\bibfnamefont{H.}~\bibnamefont{Wang}},
  \bibinfo{author}{\bibfnamefont{J.~M.} \bibnamefont{Martinis}},
  \bibnamefont{and} \bibinfo{author}{\bibfnamefont{A.~N.}
  \bibnamefont{Cleland}}, \bibinfo{journal}{Nature}
  \textbf{\bibinfo{volume}{454}}, \bibinfo{pages}{310} (\bibinfo{year}{2008}).

\bibitem[{\citenamefont{Chang et~al.}(2006)\citenamefont{Chang, S\o{}rensen,
  Hemmer, and Lukin}}]{ChangPRL06}
\bibinfo{author}{\bibfnamefont{D.~E.} \bibnamefont{Chang}},
  \bibinfo{author}{\bibfnamefont{A.~S.} \bibnamefont{S\o{}rensen}},
  \bibinfo{author}{\bibfnamefont{P.~R.} \bibnamefont{Hemmer}},
  \bibnamefont{and} \bibinfo{author}{\bibfnamefont{M.~D.} \bibnamefont{Lukin}},
  \bibinfo{journal}{Phys. Rev. Lett.} \textbf{\bibinfo{volume}{97}},
  \bibinfo{pages}{053002} (\bibinfo{year}{2006}).

\bibitem[{\citenamefont{Chang et~al.}(2007)\citenamefont{Chang, S\o{}rensen,
  Demler, and Lukin}}]{ChangNatPhy07}
\bibinfo{author}{\bibfnamefont{D.~E.} \bibnamefont{Chang}},
  \bibinfo{author}{\bibfnamefont{A.~S.} \bibnamefont{S\o{}rensen}},
  \bibinfo{author}{\bibfnamefont{E.~A.} \bibnamefont{Demler}},
  \bibnamefont{and} \bibinfo{author}{\bibfnamefont{M.~D.} \bibnamefont{Lukin}},
  \bibinfo{journal}{Nature Phys.} \textbf{\bibinfo{volume}{3}},
  \bibinfo{pages}{807} (\bibinfo{year}{2007}).

\bibitem[{\citenamefont{Liao et~al.}(2009)\citenamefont{Liao, Huang, Liu,
  Kuang, and Sun}}]{LiaoPRA09}
\bibinfo{author}{\bibfnamefont{J.-Q.} \bibnamefont{Liao}},
  \bibinfo{author}{\bibfnamefont{J.-F.} \bibnamefont{Huang}},
  \bibinfo{author}{\bibfnamefont{Y.-X.} \bibnamefont{Liu}},
  \bibinfo{author}{\bibfnamefont{L.-M.} \bibnamefont{Kuang}}, \bibnamefont{and}
  \bibinfo{author}{\bibfnamefont{C.~P.} \bibnamefont{Sun}},
  \bibinfo{journal}{Phys. Rev. A} \textbf{\bibinfo{volume}{80}},
  \bibinfo{pages}{014301} (\bibinfo{year}{2009}).

\bibitem[{\citenamefont{Zhou et~al.}(2008)\citenamefont{Zhou, Gong, Liu, Sun,
  and Nori}}]{ZhouPRL08}
\bibinfo{author}{\bibfnamefont{L.}~\bibnamefont{Zhou}},
  \bibinfo{author}{\bibfnamefont{Z.~R.} \bibnamefont{Gong}},
  \bibinfo{author}{\bibfnamefont{Y.-X.} \bibnamefont{Liu}},
  \bibinfo{author}{\bibfnamefont{C.~P.} \bibnamefont{Sun}}, \bibnamefont{and}
  \bibinfo{author}{\bibfnamefont{F.}~\bibnamefont{Nori}},
  \bibinfo{journal}{Phys. Rev. Lett.} \textbf{\bibinfo{volume}{101}},
  \bibinfo{pages}{100501} (\bibinfo{year}{2008}).

\bibitem[{\citenamefont{Zhou et~al.}(2009)\citenamefont{Zhou, Yang, Liu, Sun,
  and Nori}}]{ZhouPRA09}
\bibinfo{author}{\bibfnamefont{L.}~\bibnamefont{Zhou}},
  \bibinfo{author}{\bibfnamefont{S.}~\bibnamefont{Yang}},
  \bibinfo{author}{\bibfnamefont{Y.-x.} \bibnamefont{Liu}},
  \bibinfo{author}{\bibfnamefont{C.~P.} \bibnamefont{Sun}}, \bibnamefont{and}
  \bibinfo{author}{\bibfnamefont{F.}~\bibnamefont{Nori}},
  \bibinfo{journal}{Phys. Rev. A} \textbf{\bibinfo{volume}{80}},
  \bibinfo{pages}{062109} (\bibinfo{year}{2009}).

\bibitem[{\citenamefont{Witthaut and S\o{}rensen}(2010)}]{WitthautNJP10}
\bibinfo{author}{\bibfnamefont{D.}~\bibnamefont{Witthaut}} \bibnamefont{and}
  \bibinfo{author}{\bibfnamefont{A.~S.} \bibnamefont{S\o{}rensen}},
  \bibinfo{journal}{New J. Phys.} \textbf{\bibinfo{volume}{12}},
  \bibinfo{pages}{043052} (\bibinfo{year}{2010}).

\bibitem[{\citenamefont{Longo et~al.}(2010)\citenamefont{Longo, Schmitteckert,
  and Busch}}]{LongoPRL10}
\bibinfo{author}{\bibfnamefont{P.}~\bibnamefont{Longo}},
  \bibinfo{author}{\bibfnamefont{P.}~\bibnamefont{Schmitteckert}},
  \bibnamefont{and} \bibinfo{author}{\bibfnamefont{K.}~\bibnamefont{Busch}},
  \bibinfo{journal}{Phys. Rev. Lett.} \textbf{\bibinfo{volume}{104}},
  \bibinfo{pages}{023602} (\bibinfo{year}{2010}).

\bibitem[{\citenamefont{Vogel et~al.}(2001)\citenamefont{Vogel, Welsch, and
  Wallentowitz}}]{VogelQO03}
\bibinfo{author}{\bibfnamefont{W.}~\bibnamefont{Vogel}},
  \bibinfo{author}{\bibfnamefont{D.~G.} \bibnamefont{Welsch}},
  \bibnamefont{and}
  \bibinfo{author}{\bibfnamefont{S.}~\bibnamefont{Wallentowitz}},
  \emph{\bibinfo{title}{Quantum Optics-An Introduction}}
  (\bibinfo{publisher}{WILEY-VCH}, \bibinfo{address}{Berlin},
  \bibinfo{year}{2001}), \bibinfo{edition}{2nd} ed.

\bibitem[{\citenamefont{Thompson et~al.}(1992)\citenamefont{Thompson, Rempe,
  and Kimble}}]{ThompsonPRL92}
\bibinfo{author}{\bibfnamefont{R.~J.} \bibnamefont{Thompson}},
  \bibinfo{author}{\bibfnamefont{G.}~\bibnamefont{Rempe}}, \bibnamefont{and}
  \bibinfo{author}{\bibfnamefont{H.~J.} \bibnamefont{Kimble}},
  \bibinfo{journal}{Phys. Rev. Lett.} \textbf{\bibinfo{volume}{68}},
  \bibinfo{pages}{1132} (\bibinfo{year}{1992}).

\bibitem[{\citenamefont{Reithmaier et~al.}(2004)\citenamefont{Reithmaier, Sek,
  L\"offler, Hofmann, Kuhn, Reitzenstein, Keldysh, Kulakovskii, Reinecke, and
  Forchel}}]{ReithmaierNat04}
\bibinfo{author}{\bibfnamefont{J.~P.} \bibnamefont{Reithmaier}},
  \bibinfo{author}{\bibfnamefont{G.}~\bibnamefont{Sek}},
  \bibinfo{author}{\bibfnamefont{A.}~\bibnamefont{L\"offler}},
  \bibinfo{author}{\bibfnamefont{C.}~\bibnamefont{Hofmann}},
  \bibinfo{author}{\bibfnamefont{S.}~\bibnamefont{Kuhn}},
  \bibinfo{author}{\bibfnamefont{S.}~\bibnamefont{Reitzenstein}},
  \bibinfo{author}{\bibfnamefont{L.~V.} \bibnamefont{Keldysh}},
  \bibinfo{author}{\bibfnamefont{V.~D.} \bibnamefont{Kulakovskii}},
  \bibinfo{author}{\bibfnamefont{T.~L.} \bibnamefont{Reinecke}},
  \bibnamefont{and} \bibinfo{author}{\bibfnamefont{A.}~\bibnamefont{Forchel}},
  \bibinfo{journal}{Nature} \textbf{\bibinfo{volume}{432}},
  \bibinfo{pages}{197} (\bibinfo{year}{2004}).

\bibitem[{\citenamefont{Wallraff et~al.}(2004)\citenamefont{Wallraff, Schuster,
  Blais, Frunzio, Huang, Majer, Kumar, Girvin, and Schoelkopf}}]{WallraffNat04}
\bibinfo{author}{\bibfnamefont{A.}~\bibnamefont{Wallraff}},
  \bibinfo{author}{\bibfnamefont{D.~I.} \bibnamefont{Schuster}},
  \bibinfo{author}{\bibfnamefont{A.}~\bibnamefont{Blais}},
  \bibinfo{author}{\bibfnamefont{L.}~\bibnamefont{Frunzio}},
  \bibinfo{author}{\bibfnamefont{R.-S.} \bibnamefont{Huang}},
  \bibinfo{author}{\bibfnamefont{J.}~\bibnamefont{Majer}},
  \bibinfo{author}{\bibfnamefont{S.}~\bibnamefont{Kumar}},
  \bibinfo{author}{\bibfnamefont{S.~M.} \bibnamefont{Girvin}},
  \bibnamefont{and} \bibinfo{author}{\bibfnamefont{R.~J.}
  \bibnamefont{Schoelkopf}}, \bibinfo{journal}{Nature}
  \textbf{\bibinfo{volume}{431}}, \bibinfo{pages}{162} (\bibinfo{year}{2004}).

\bibitem[{\citenamefont{Chiorescu et~al.}(2004)\citenamefont{Chiorescu, Bertet,
  Semba, Nakamura, Harmans, and Mooij}}]{ChiorescuNat04}
\bibinfo{author}{\bibfnamefont{I.}~\bibnamefont{Chiorescu}},
  \bibinfo{author}{\bibfnamefont{P.}~\bibnamefont{Bertet}},
  \bibinfo{author}{\bibfnamefont{K.}~\bibnamefont{Semba}},
  \bibinfo{author}{\bibfnamefont{Y.}~\bibnamefont{Nakamura}},
  \bibinfo{author}{\bibfnamefont{C.~J. P.~M.} \bibnamefont{Harmans}},
  \bibnamefont{and} \bibinfo{author}{\bibfnamefont{J.~E.} \bibnamefont{Mooij}},
  \bibinfo{journal}{Nature} \textbf{\bibinfo{volume}{431}},
  \bibinfo{pages}{159} (\bibinfo{year}{2004}).

\bibitem[{\citenamefont{Astafiev et~al.}(2010)\citenamefont{Astafiev, Zagoskin,
  Abdumalikov, Pashkin, Yamamoto, Inomata, Nakamura, and Tsai}}]{AstafievSci10}
\bibinfo{author}{\bibfnamefont{O.}~\bibnamefont{Astafiev}},
  \bibinfo{author}{\bibfnamefont{A.~M.} \bibnamefont{Zagoskin}},
  \bibinfo{author}{\bibfnamefont{A.~A.~J.} \bibnamefont{Abdumalikov}},
  \bibinfo{author}{\bibfnamefont{Y.~A.} \bibnamefont{Pashkin}},
  \bibinfo{author}{\bibfnamefont{T.}~\bibnamefont{Yamamoto}},
  \bibinfo{author}{\bibfnamefont{K.}~\bibnamefont{Inomata}},
  \bibinfo{author}{\bibfnamefont{Y.}~\bibnamefont{Nakamura}}, \bibnamefont{and}
  \bibinfo{author}{\bibfnamefont{J.~S.} \bibnamefont{Tsai}},
  \bibinfo{journal}{Science} \textbf{\bibinfo{volume}{327}},
  \bibinfo{pages}{840} (\bibinfo{year}{2010}).

\bibitem[{\citenamefont{Schoelkopf and Girvin}(2008)}]{SchoelkopfNat08}
\bibinfo{author}{\bibfnamefont{R.~J.} \bibnamefont{Schoelkopf}}
  \bibnamefont{and} \bibinfo{author}{\bibfnamefont{S.~M.}
  \bibnamefont{Girvin}}, \bibinfo{journal}{Nature}
  \textbf{\bibinfo{volume}{451}}, \bibinfo{pages}{664} (\bibinfo{year}{2008}).

\bibitem[{\citenamefont{Longo et~al.}(2009)\citenamefont{Longo, Schmittechert,
  and Busch}}]{LongoJOA09}
\bibinfo{author}{\bibfnamefont{P.}~\bibnamefont{Longo}},
  \bibinfo{author}{\bibfnamefont{P.}~\bibnamefont{Schmittechert}},
  \bibnamefont{and} \bibinfo{author}{\bibfnamefont{K.}~\bibnamefont{Busch}},
  \bibinfo{journal}{J. Opt. A} \textbf{\bibinfo{volume}{11}},
  \bibinfo{pages}{114009} (\bibinfo{year}{2009}).

\bibitem[{\citenamefont{Shen and Fan}(2007{\natexlab{a}})}]{ShenPRL07}
\bibinfo{author}{\bibfnamefont{J.-T.} \bibnamefont{Shen}} \bibnamefont{and}
  \bibinfo{author}{\bibfnamefont{S.}~\bibnamefont{Fan}},
  \bibinfo{journal}{Phys. Rev. Lett.} \textbf{\bibinfo{volume}{98}},
  \bibinfo{pages}{153003} (\bibinfo{year}{2007}{\natexlab{a}}).

\bibitem[{\citenamefont{Shen and Fan}(2007{\natexlab{b}})}]{ShenPRA07}
\bibinfo{author}{\bibfnamefont{J.-T.} \bibnamefont{Shen}} \bibnamefont{and}
  \bibinfo{author}{\bibfnamefont{S.}~\bibnamefont{Fan}},
  \bibinfo{journal}{Phys. Rev. A} \textbf{\bibinfo{volume}{76}},
  \bibinfo{pages}{062709} (\bibinfo{year}{2007}{\natexlab{b}}).

\bibitem[{\citenamefont{Maunz et~al.}(2007)\citenamefont{Maunz, Moehring,
  Olmschenk, Younge, Matsukevich, and Monroe}}]{MaunzNatPhys07}
\bibinfo{author}{\bibfnamefont{P.}~\bibnamefont{Maunz}},
  \bibinfo{author}{\bibfnamefont{D.~L.} \bibnamefont{Moehring}},
  \bibinfo{author}{\bibfnamefont{S.}~\bibnamefont{Olmschenk}},
  \bibinfo{author}{\bibfnamefont{K.~C.} \bibnamefont{Younge}},
  \bibinfo{author}{\bibfnamefont{D.~N.} \bibnamefont{Matsukevich}},
  \bibnamefont{and} \bibinfo{author}{\bibfnamefont{C.}~\bibnamefont{Monroe}},
  \bibinfo{journal}{Nature Phys.} \textbf{\bibinfo{volume}{3}},
  \bibinfo{pages}{538} (\bibinfo{year}{2007}).

\bibitem[{\citenamefont{Nishino et~al.}(2009)\citenamefont{Nishino, Imamura,
  and Hatano}}]{NishinoPRL09}
\bibinfo{author}{\bibfnamefont{A.}~\bibnamefont{Nishino}},
  \bibinfo{author}{\bibfnamefont{T.}~\bibnamefont{Imamura}}, \bibnamefont{and}
  \bibinfo{author}{\bibfnamefont{N.}~\bibnamefont{Hatano}},
  \bibinfo{journal}{Phys. Rev. Lett.} \textbf{\bibinfo{volume}{102}},
  \bibinfo{pages}{146803} (\bibinfo{year}{2009}).

\bibitem[{\citenamefont{Imamura et~al.}(2009)\citenamefont{Imamura, Nishino,
  and Hatano}}]{ImamuraPRB09}
\bibinfo{author}{\bibfnamefont{T.}~\bibnamefont{Imamura}},
  \bibinfo{author}{\bibfnamefont{A.}~\bibnamefont{Nishino}}, \bibnamefont{and}
  \bibinfo{author}{\bibfnamefont{N.}~\bibnamefont{Hatano}},
  \bibinfo{journal}{Phys. Rev. B} \textbf{\bibinfo{volume}{80}},
  \bibinfo{pages}{245323} (\bibinfo{year}{2009}).

\bibitem[{\citenamefont{Sakurai}(1994)}]{SakuraiQM94}
\bibinfo{author}{\bibfnamefont{J.~J.} \bibnamefont{Sakurai}},
  \emph{\bibinfo{title}{Modern Quantum Mechanics}}
  (\bibinfo{publisher}{Addison-Wesley}, \bibinfo{address}{Reading, MA},
  \bibinfo{year}{1994}).

\bibitem[{\citenamefont{Loudon}(2003)}]{LoudonQTL03}
\bibinfo{author}{\bibfnamefont{R.}~\bibnamefont{Loudon}},
  \emph{\bibinfo{title}{The Quantum Theory of Light}}
  (\bibinfo{publisher}{Oxford University Press}, \bibinfo{address}{New York},
  \bibinfo{year}{2003}), \bibinfo{edition}{3rd} ed.

\bibitem[{\citenamefont{Hwang}(2003)}]{HwangPRL03}
\bibinfo{author}{\bibfnamefont{W.-Y.} \bibnamefont{Hwang}},
  \bibinfo{journal}{Phys. Rev. Lett.} \textbf{\bibinfo{volume}{91}},
  \bibinfo{pages}{057901} (\bibinfo{year}{2003}).

\bibitem[{\citenamefont{Lo et~al.}(2005)\citenamefont{Lo, Ma, and
  Chen}}]{LoPRL05}
\bibinfo{author}{\bibfnamefont{H.-K.} \bibnamefont{Lo}},
  \bibinfo{author}{\bibfnamefont{X.}~\bibnamefont{Ma}}, \bibnamefont{and}
  \bibinfo{author}{\bibfnamefont{K.}~\bibnamefont{Chen}},
  \bibinfo{journal}{Phys. Rev. Lett.} \textbf{\bibinfo{volume}{94}},
  \bibinfo{pages}{230504} (\bibinfo{year}{2005}).

\bibitem[{\citenamefont{Wang}(2005)}]{WangPRL05}
\bibinfo{author}{\bibfnamefont{X.-B.} \bibnamefont{Wang}},
  \bibinfo{journal}{Phys. Rev. Lett.} \textbf{\bibinfo{volume}{94}},
  \bibinfo{pages}{230503} (\bibinfo{year}{2005}).

\bibitem[{\citenamefont{Curty et~al.}(2009)\citenamefont{Curty, Moroder, Ma,
  and L\"utkenhaus}}]{MCOPTLETT09}
\bibinfo{author}{\bibfnamefont{M.}~\bibnamefont{Curty}},
  \bibinfo{author}{\bibfnamefont{T.}~\bibnamefont{Moroder}},
  \bibinfo{author}{\bibfnamefont{X.}~\bibnamefont{Ma}}, \bibnamefont{and}
  \bibinfo{author}{\bibfnamefont{N.}~\bibnamefont{L\"utkenhaus}},
  \bibinfo{journal}{Opt. Lett.} \textbf{\bibinfo{volume}{34}},
  \bibinfo{pages}{3238} (\bibinfo{year}{2009}).

\end{thebibliography}
\end{document}